\documentstyle[twocolumn,seceq,bm,xspace,graphicx]{jpsj}

\newcommand{\urrs}{U(Ru$_{1-x}$Rh$_x$)$_2$Si$_2$\xspace }

\newcommand{\urs}{URu$_2$Si$_2$\xspace }
\newcommand{\tord}{$T_{\rm o}$\xspace}

\newcommand{\tmag}{$T_{\rm m}$\xspace}
\newcommand{\tMag}{$T_{\rm M}$\xspace}
\newcommand{\moment}{$\mu_{\rm o}$\xspace}

\newcommand{\mb}{$\mu_{\rm B}/{\rm U}$\xspace}

\def\runtitle{Neutron Scattering Study on Competition between Hidden Order and Antiferromagnetism in U(Ru$_{1-x}$Rh$_x$)$_2$Si$_2$ ($x \le 0.05$)}
\def\runauthor{Makoto {\sc Yokoyama} {\it et al}.}

\title
{
Neutron Scattering Study on Competition between Hidden Order and Antiferromagnetism in U(Ru$_{\bm{1-x}}$Rh$_{\bm{x}}$)$_{\bm 2}$Si$_{\bm 2}$ ($\bm{x \le 0.05}$)
}

\author
{ 
Makoto {\sc Yokoyama}\footnote{Email: makotti@mx.ibaraki.ac.jp}, Hiroshi {\sc Amitsuka},$^1$ Seiichiroh {\sc Itoh},$^1$ Ikuto {\sc Kawasaki},$^1$\\ Kenichi {\sc Tenya}$^1$ and  Hideki {\sc Yoshizawa}$^2$
}

\inst
{
Faculty of Science, Ibaraki University, Mito 310-8512\\
$^1$Graduate School of Science, Hokkaido University, Sapporo 060-0810\\
$^2$Neutron Science Laboratory, Institute for Solid State Physics, The University of Tokyo, Tokai 319-1106\\
}

\recdate
{
\today
}

\abst
{
We have performed elastic and inelastic neutron scattering experiments on the solid solutions U(Ru$_{1-x}$Rh$_x$)$_2$Si$_2$ for the Ru rich concentrations: $x=0,\ 0.01,\ 0.02,\ 0.025,\ 0.03,\ 0.04$ and 0.05. Hidden order is suppressed with increasing $x$, and correspondingly the onset temperature \tmag ($\sim$ 17.5 K at $x=0$) of weak antiferromagnetic (AF) Bragg reflection decreases. For $x=0.04$ and 0.05, no magnetic order is detected in the investigated temperature range down to 1.4 K. In the middle range, $0.02 \le x \le 0.03$, we found that the AF Bragg reflection is strongly enhanced. At $x=0.02$, this takes place at $\sim$ 7.7 K ($\equiv T_{\rm M}$), which is significantly lower than \tmag ($\sim$ 13.7 K). \tMag increases with increasing $x$, and seems to merge with \tmag at $x=0.03$. If the AF state is assumed to be homogeneous, the staggered moment \moment estimated at 1.4 K increases from 0.02(2) \mb ($x=0$) to 0.24(1) \mb ($x=0.02$). The behavior is similar to that observed under hydrostatic pressure (\moment increases to $\sim$ 0.25 \mb at 1.0 GPa), suggesting that the AF evolution induced by Rh doping is due to an increase in the AF volume fraction. We also found that the magnetic excitation observed at $Q=(1,0,0)$ below \tmag disappears as $T$ is lowered below \tMag .
}

\kword
{
URu$_2$Si$_2$, hidden order, inhomogeneous magnetism, Rh substitution, neutron scattering
}

\begin{document}
\sloppy
\maketitle
The relation between the phase transition at $T_{\rm o}=17.5\ {\rm K}$ and the weak antiferromagnetsm in the heavy-electron compound \urs (the ThCr$_2$Si$_2$ type, body-centered tetragonal structure)\cite{rf:Palstra85,rf:Schlabitz86,rf:Maple86} has been intensively investigated since the finding of unusually small antiferromagnetic (AF) moment below $\sim$\tord.\cite{rf:Broholm87,rf:Mason90,rf:Fak96} Recent microscopic investigations performed under hydrostatic pressure $P$ shed a new light on this issue.\cite{rf:Ami99,rf:Matsuda2001} The high-$P$ study using neutron scattering revealed that the AF Bragg-peak intensities are strongly enhanced by applying $P$. If the AF order is assumed to be homogeneous, the estimated staggered moment \moment continuously increases from 0.02 \mb ($P=0$) to 0.25 \mb (1.0 GPa).\cite{rf:Ami99} In parallel, the $^{29}$Si-NMR measurements revealed that the volume fraction of AF order develops inhomogeneously with increasing $P$.\cite{rf:Matsuda2001}  In the AF-order region, the magnitude of the internal fields is nearly independent of $P$. The $P$ dependence of the AF volume fraction is roughly in proportion to $\mu_{\rm o}^2(P)$ derived from neutron scattering, indicating that the observed enhancement of the AF Bragg reflection is attributed to the AF volume expansion. A simple extrapolation yields the AF volume fraction at ambient pressure of $\sim$ 1\%, and we believe this to be the true nature of the small AF moment. The remaining 99\% of the system is thus considered to be occupied by unidentified ``hidden order" (HO), which is responsible for the large bulk anomalies observed at \tord , such as a specific-heat jump ($\Delta C/T_{\rm o}\sim 300\ {\rm mJ/K^2 mol}$).

Quite recently, we performed neutron scattering experiments under uniaxial stress $\sigma$.\cite{rf:Yoko2002,rf:Yoko2003} We found that the AF Bragg-peak intensity is strongly enlarged by applying $\sigma$ along the tetragonal basal plane. The $\mu_{\rm o}$ values obtained for $\sigma$\,$\parallel$\,[100] and [110] continuously increase from $\sim$ 0.02 \mb ($\sigma=0$) to $\sim$ 0.22 \mb (0.25 GPa), showing a close resemblance to $\mu_{\rm o}(P)$. On the basis of a crystal-strain model, we pointed out that the $\mu_{\rm o}(\sigma)$ and $\mu_{\rm o}(P)$ data can fairly well be scaled by the $c/a$ ratio ($\eta$), and suggested that $\eta$ is an intrinsic parameter to yield the competition between HO and the AF order. The analyses also predict that a slight increase in $\eta$ of $\sim$ 0.1\% may induces the AF order of nearly the full volume fraction.

The axial type lattice distortion may be expected not only by compression but also by alloying. In the Rh-substitution systems \urrs , HO phase is known to be suppressed at $x\sim 0.04$, and then replaced by a new AF state with a complex multi-$Q$ structure in the range between $x\sim 0.1$ and 0.3.\cite{rf:Ami88,rf:Dalichaouch90,rf:Miyako91,rf:Burlet92,rf:Kawa94} The X-ray powder diffraction measurements\cite{rf:Burlet92} revealed that $\eta$ linearly increases with increasing $x$ at a rate: $\partial \ln\eta /\partial x \sim 7\times 10^{-2}$, indicating that the 0.1\% increase of $\eta$ is achieved at $x\sim 0.02$. It is thus expected that the inhomogeneous AF state is significantly induced by small amount of Rh-doping. In order to see this possibility, we have investigated microscopic properties on \urrs ($x\le 0.05$) by performing neutron scattering experiments.

Single-crystalline samples \urrs for $x=0$, 0.01, 0.02, 0.025, 0.03, 0.04 and 0.05 were grown by the Czochralski pulling method using a tetra-arc furnace, and vacuum-annealed at 1000$^\circ$C for 5 days. The cube-shaped samples were cut out of the ingots by means of spark erosion, mounted in aluminum cans filled with $^4$He gas, and cooled down to 1.4 K in a $^4$He cryostat.  The elastic and inelastic neutron scattering measurements were performed on the triple-axis spectrometer GPTAS (4G) located at the JRR-3M research reactor of Japan Atomic Energy Research Institute. We chose the ($hk0$) scattering plane in both the measurements. For elastic scattering, the neutron momentum $k=2.66\ {\rm \AA^{-1}}$ was selected by using the (002) reflection of pyrolytic graphite (PG) for both the monochromator and the analyzer. We used the combination of 40'-80'-40'-80' horizontal collimators, together with two PG filters to eliminate the higher order reflections. The scans were performed on the AF Bragg reflections from (100) and (210), and the nuclear ones from (200), (110) and (020). For inelastic scattering, we made constant-$Q$ scans at $Q=(1,0,0)$ using neutrons with a fixed final momentum $k_{\rm f}=2.65\ {\rm \AA^{-1}}$. The combination of 40'-80'-40'-80' collimators and one PG filter was chosen. The energy resolution determined from the full width of the half maximum of the vanadium incoherent peak was $\sim$ 0.88 meV.

Figure 1 shows the $x$ variations of the AF Bragg-peak profiles at 1.4 K, obtained from the longitudinal scans at (100). Instrumental backgrounds and higher-order nuclear reflections were carefully subtracted by using the data taken at 40 K. The (100) AF Bragg-peak intensity increases about 1.5 times with increasing $x$ from 0 to 0.01, and then abruptly develops about hundred times at $x=0.02$, followed by a gradual decrease at $x=0.025$ and 0.03. No Bragg reflection was observed down to 1.4 K for $x=0.04$ and 0.05. 
\begin{figure}[tbp]
\begin{center}
\includegraphics[keepaspectratio,width=0.48\textwidth]{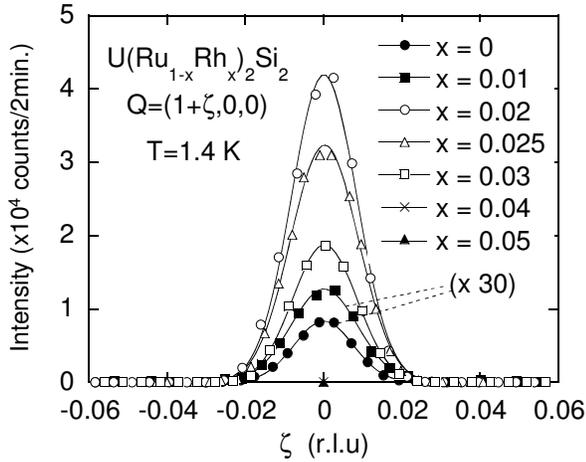}
\end{center}
\caption{The magnetic Bragg peaks of \urrs ($x\le 0.5$) at 1.4 K, obtained from the longitudinal scans at the (100) position. Note that the intensities for $x=0$ and 0.01 are 30 times enlarged.}
\end{figure}

Similar $x$ dependence of the AF Bragg peak was observed for the scans at (210), and for each Rh concentration the integrated intensities of the (100) and (210) peaks roughly follow the $|Q|$ dependence expected from the U$^{4+}$ magnetic form factor\cite{rf:Frazer65} by taking polarization factor unity. In addition, we detected no other Bragg reflection in the scans along the principal axes of the Brillouin Zone: ($1+\zeta$,0,0) for $0\le \zeta \le 1$ and ($2-\zeta$,$\zeta$,0) for $0\le \zeta \le 0.5$. These results indicate that the type-I AF structure with the uranium 5f moment polarized along the $c$ axis is unchanged by doping Rh in the range $x\le 0.03$.

Displayed in Fig.\ 2 is the temperature variations of the integrated intensity $I(T)$ of the (100) magnetic reflection for $x\le 0.04$. For $x=0$ and 0.01, $I(T)$ shows an unusually slow development with decreasing temperature. The onset of this weak reflection \tmag slightly decreases from $\sim$ 17.5 K ($x=0$) to $\sim$15.9 K ($x=0.01$). In contrast, in the range $0.02\le x \le 0.03$ $I(T)$ exhibits a dramatic $T$ variation. At $x=0.02$, the magnetic reflection slightly develops below $T_{\rm m}\sim 13.7\ {\rm K}$, and shows an abrupt increase at $\sim$ 7.7 K ($\equiv T_{\rm M}$). Below $\sim$ 4 K $I(T)$ slightly decreases. The increase of $I(T)$ at \tMag is sharper than that expected from the typical second-order phase transition. We carefully took the data points with decreasing and increasing $T$, but detected no hysteresis around \tMag and \tmag within the experimental accuracy. Similar $I(T)$ curves are obtained for $x=0.025$ and 0.03, where the interval between \tmag and \tMag becomes narrower than in $x=0.02$. For $x=0.04$, on the other hand, no anomaly is found at least in the investigated range between 1.4 K and 40 K.
\begin{figure}[tbp]
\begin{center}
\includegraphics[keepaspectratio,width=0.48\textwidth]{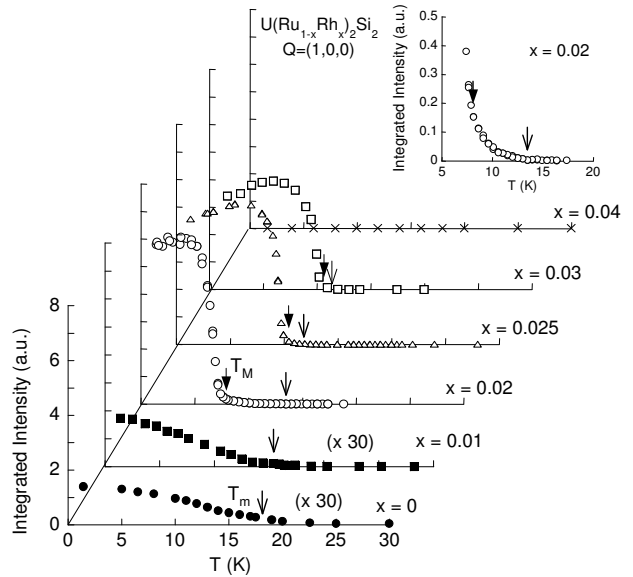}
\end{center}
\caption{Temperature variations of the integrated intensity for the (100) magnetic reflections of \urrs ($x \le 0.04$). The enlarged figure for $x=0.02$ between 5 K and 20 K is shown in the inset. The arrows indicate the onset temperatures \tmag and \tMag defined in the text.}
\end{figure}

In Fig.\ 3, we plot the characteristic temperatures \tord , \tmag and \tMag, and the staggered moment \moment as a function of $x$, where the hidden-order transition temperature \tord is obtained from the specific-heat measurements.\cite{rf:Ami2004} The \moment values are estimated from the integrated intensities of the (100) magnetic Bragg peaks, normalized by the intensities of the weak (110) nuclear reflection. Note that the present estimation of \moment is based on the assumption that the AF order is homogeneous. For pure \urs , the magnitude of \moment at 1.4 K is estimated to be 0.02(2) \mb , which is in good agreement with the values of the previous reports.\cite{rf:Broholm87,rf:Mason90,rf:Fak96,rf:Ami99,rf:Yoko2002} By substituting Rh for Ru, \moment still stays in the same order as in $x=0$, in the $x-T$ range of $T<T_{\rm m}$ ($x\le 0.01$) and $T_{\rm M}<T<T_{\rm m}$ ($0.02\le x \le 0.03$). This result, together with the fact that \tord $\sim$ \tmag ($x\le 0.03$), indicates that the HO occupies the system as the majority phase that is competing with a small amount of the inhomogeneous AF phase in the $x-T$ region $T_{\rm M}<T<T_{\rm m}$ ($x\le 0.03$). On the other hand, for $x=0.02$ \moment increases to 0.24(1) \mb ($T=1.4\ {\rm K}$), which roughly corresponds to the values ($\sim$ 0.25 and 0.22 {\mb}) observed in the AF majority phase of \urs under $P$ ($\sim$ 1.0 GPa) and $\sigma$\,$\bot$\,[001] ($\sim$ 0.25 GPa), respectively. This suggests that the development of \moment caused by Rh substitution below \tMag is also due to the increase in the volume fraction of the AF phase. Our recent zero-field $\mu$SR measurements for \urrs ($x\le 0.04$)\cite{rf:Ami2004} support this suggestion, where the AF volume fraction below \tMag for $x=0.02$ is estimated to be about 70\%.
\begin{figure}[tbp]
\begin{center}
\includegraphics[keepaspectratio,width=0.48\textwidth]{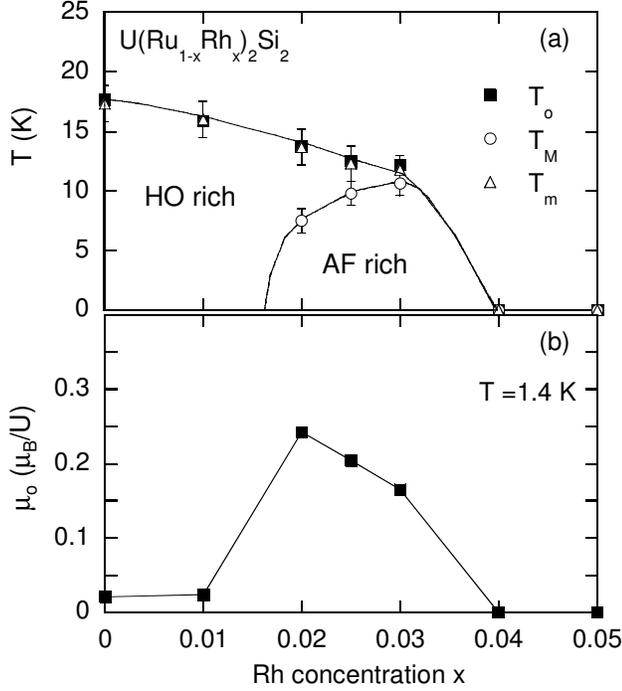}
\end{center}
\caption{(a) The transition temperature \tord , the onset temperatures of the magnetic Bragg peak \tmag and \tMag , and (b) the staggered moment \moment as a function of Rh concentration $x$. The lines are guides to the eye.}
\end{figure}

As $x$ increases above 0.02, the values of \tord, \tmag and \moment decrease, and then vanish at $x=0.04$. This behavior is in contrast with that observed in pure \urs under $P$ and $\sigma$, where those quantities increase with $P$ and $\sigma$. \cite{rf:Ami99,rf:Yoko2002,rf:Yoko2003,rf:Louis86,rf:McElfresh87,rf:Fisher90,rf:Ido93,rf:Oomi94,rf:Bakker92} This difference may be explained as that the AF order becomes more stable than HO in the range $0.02\le x \le 0.03$ mainly because of the $c/a$ extension, but both of them are suppressed by some other effect which is characteristic of the Rh substitution, such as a disorder of the crystal periodicity and a variation of the conduction electron number.

In the inelastic scattering experiments for $x=0.02$, we used a different single crystal from that used in the elastic scattering experiments. We checked by the elastic scans that $I(T)$ of this sample shows almost the same $T$ dependence as that of another one (Fig.\ 2) except a slightly higher \tMag ($\sim$ 8.3 K), which will be ascribed to an error ($\sim$ 10\%) of $x$. In Fig.\ 4, we plot the temperature variations of the energy scans at (100) for $x=0.02$, after subtracting the contributions of the incoherent scattering and the instrumental background. The instrumental background was measured by scans at the corresponding $|Q|$-invariant position (0.707,0.707,0), where we observed neither magnon nor phonon excitations. The sharp peak at the energy transfer $\hbar\omega\sim 0$ arises mainly from the higher-order nuclear Bragg reflections and the AF Bragg reflections. We found a broad peak anomaly due to magnetic excitations appearing below $\sim$ 15 K, and growing significantly below $\sim$ \tord ($=13.7\ {\rm K}$). Although it is impossible to determine the peak position within the present experimental accuracy, it is roughly estimated to be around 0.5 meV or lower, which is significantly smaller than that for pure \urs ($\sim$ 2.4 meV).\cite{rf:Broholm87,rf:Ami99} Interestingly, as temperature is lowered below \tMag , this peak suddenly disappears. For $1.4\ {\rm K}\le T\le T_{\rm M}$, we could observe no anomaly within the energy range of $\hbar\omega \le 10\ {\rm meV}$. 
\begin{figure}[tbp]
\begin{center}
\includegraphics[keepaspectratio,width=0.48\textwidth]{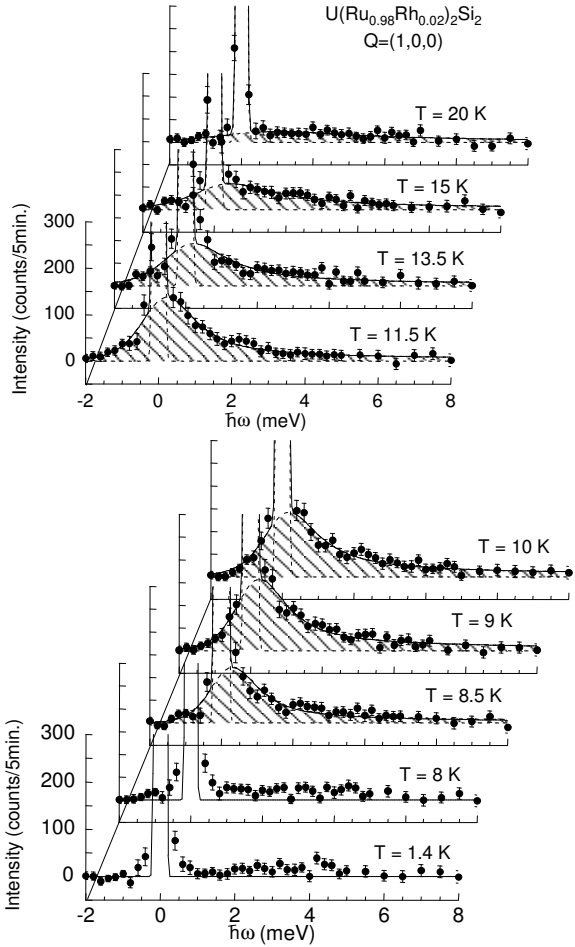}
\end{center}
\caption{Temperature variations of constant-$Q$ scans for $Q=(100)$ in U(Ru$_{0.98}$Rh$_{0.02}$)$_2$Si$_2$. Incoherent scattering and instrumental background are subtracted. The lines and the shades are guides to the eye.}
\end{figure}

All of these experimental results are reasonably well compared with the features obtained by the previous high-$P$ and $\sigma$ measurements on the pure compound\cite{rf:Ami99,rf:Matsuda2001,rf:Yoko2002,rf:Yoko2003,rf:Ami2003-1,rf:Ami2003-2,rf:Motoyama2003}: (i) HO and AF phases are competing with each other via a first-order phase transition, (ii) the $c/a$ ratio ($\eta$) governs the switching of the two phases, and (iii) magnetic excitations at $Q= (1,0,0)$ in HO vanish in the AF phase. 

Regarding the features (i) and (ii), the Rh doping induces the AF phase at $x\sim 0.02$, where $\eta$ is estimated to be $\sim$ 0.1\% in agreement with the value predicted by the high-$P$ ($\sigma$) studies. One remarkable difference would be the abrupt increase seen in $I(T)$ at \tMag , which is clearly lower than \tord . This behavior is not indicated in the previous neutron-scattering\cite{rf:Ami99,rf:Yoko2002,rf:Yoko2003} and NMR\cite{rf:Matsuda2001} data, where the AF volume fraction continuously increases below $\sim$ \tord . On the other hand, recent thermal-expansion\cite{rf:Motoyama2003} and $\mu$SR\cite{rf:Ami2003-1} measurements provide a similar phase diagram to the present one, exhibiting a double transition. We consider that this difference mostly arises from a distribution of local strains in the crystal. Because of the smallness in the critical $\eta$ value ($\sim$ 0.1\%), a slight difference in the distribution of local $\eta$ values would cause a significant effect on the behavior of AF evolution. If the distribution is small, a sharp onset of the AF phase will be expected at a temperature below \tord and at finite values of $P$, $\sigma$ or $x$; while if it is broad, the phase boundary will be smeared.\cite{rf:Ami2003-2} The observed abrupt increase of the AF phase with $T$ and $x$ thus indicates that the Rh doping does not strongly affect the distribution of $\eta$. 

One could argue that the absence of hysteresis in $I(T)$ near \tMag ($\sim$ 8 K) is inconsistent with the first-order phase transition. This is however understood if the potential barrier between the two types order is small. We observed a clear hysteresis loop of $I(T)$ in a $\sigma$ sweep performed at 1.4 K.\cite{rf:Yoko2002} The barrier height is thus expected to be between these two temperatures, say roughly a few Kelvins in energy.

We should also note that the obtained $x-T$ phase diagram indicates the presence of a bicritical point at $x\sim 0.03$, where \tMag and \tord seem to merge with each other. This supports the argument that the free energy of this system has a coupling term of the type $\psi^2m^2$.\cite{rf:Shah2000,rf:Chandra2002} In this situation, it is not necessary for the hidden order parameter $\psi$ to break time reversal symmetry and have the same ordering $Q$ vector as the dipole moment $m$. Such possibilities have actually been argued in the models that assume $\psi$ to be quadrupoles, \cite{rf:Ami94,rf:Santini94, rf:Ohkawa99,rf:Tsuruta2000} bond-currents \cite{rf:Chandra2002} and uranium dimers. \cite{rf:Kasuya97} We wish to stress that among them the quadrupole order with $\psi = J_x^2-J_y^2$ or $J_xJ_y + J_yJ_x$ may give a simplest explanation for the feature (iii), as $\psi$ and $m$ are orthogonal and can be represented as $S_{x,y}$ and $S_z$ of the pseudo spin $S = 1/2$.\cite{rf:Ohkawa99}

In conclusion, the mixed compounds \urrs ($x \le 0.05$) provide another example by which one can observe (and tune) the competition between HO and the AF order in \urs . Except the suppression of both the phases at $x\sim 0.04$, the major features are consistent with those obtained by previous measurements performed under hydrostatic pressure and uniaxial stress. Because of no restriction by pressure cells, it is expected that further insights of this unusual competition will be obtainable from the detailed studies using this system.

We are grateful to J.A.\ Mydosh for helpful discussions. This work was supported by a Grant-In-Aid for Scientific Research from Ministry of Education, Sport, Science and Culture of Japan.

\end{document}